\documentclass[twocolumn,showpacs,amsmath,amssymb,superscriptaddress,aps,prc]{revtex4-1}
\usepackage[dvips]{graphicx}
\usepackage{epsfig}
\usepackage{amsmath}
\usepackage[dvips]{color}
\newcommand{\vis}[1]{\mbox{\boldmath $#1$}}

\begin{document}
\title{Microscopic description of $^7$Li in the $^{7}\text{Li}+^{12}\text{C}$ and $^7\text{Li}+^{28}\text{Si}$ elastic scattering at high energies}
\author{E. C. Pinilla}
\author{P. Descouvemont}
\affiliation{Physique Nucl\'eaire Th\'eorique et Physique Math\'ematique, C.P. 229,\\
Universit\'e Libre de Bruxelles (ULB), B 1050 Brussels, Belgium}
\date{\today}
\begin{abstract}
We employ a microscopic continuum-discretized coupled-channels reaction framework (MCDCC) to study the elastic angular distribution of the $^7$Li$=\alpha+t$ nucleus colliding with $^{12}$C and $^{28}$Si targets at $E_{\text{Lab}}$=350 MeV. In this framework, the $^7$Li projectile is described in a microscopic cluster model and impinges on non-composite targets. The diagonal and coupling potentials are constructed from nucleon-target interactions and $^7$Li microscopic wave functions. We obtain a fair description of the experimental data, in the whole angular range studied, when continuum channels are included. The inelastic and breakup angular distributions on the lightest target are also investigated. In addition, we compute $^{7}$Li$+^{12}$C MCDCC elastic cross sections at energies much higher than the Coulomb barrier and we use them as reference calculations to test the validity of multichannel eikonal cross sections. 
\end{abstract}
\maketitle
\section{Introduction}
Exotic nuclei are at the limit of the stability lines and exhibit unusual properties, such as a large radius \cite{THH85}. The specific properties of these nuclei must be included in the wave function in order to compare reaction theories with experiments. Then, a reliable description of a reaction process involving exotic nuclei, must combine an accurate projectile wave function and an appropriate reaction model. Light exotic nuclei are known to group in substructures with its own identity or clusters. Typical examples are the $^7$Li nucleus seen as made of $\alpha$ and $t$ substructures, and the two neutron halo nuclei $^6$He and $^{11}$Li, seen as $\alpha$ and $^{9}$Li cores plus two neutrons. In order to describe the structure of such nuclei, microscopic \cite{WT77,SV98,Ka86} cluster models and their few-body approximations \cite{ZDF93,DTV98} have been implemented.

Few-body approximations of microscopic cluster models are built on nucleus-nucleus or nucleus-nucleon interactions and the Pauli principle between clusters is simulated by a suitable choice of those interactions \cite{Ba87,TDE00,KP78}. Even though they are easier to interpret and to integrate in reaction models (see for instance \cite{GBC06,BCD09,ACJ07}), they present some drawbacks as: i) the required nucleus-nucleus potentials are generally poorly known or not known at all. ii) Inaccuracy introduced by considering the Pauli principle approximately \cite{PBD11}. iii) In most of the calculations, core excitations are neglected. In contrast, microscopic cluster models are based on nucleon-nucleon interactions. Hence they are expected to be more precise. Their main advantages are: i) they take exactly the Pauli principle into account. ii) Core excitations can be included in a direct way. Therefore, a significant improvement of current reaction calculations in exotic nuclei should contain a microscopic description of the projectile.

For weakly bound nuclei, we expect that continuum states influence most of the reaction processes. At low energies, around the Coulomb barrier, we can study this influence within the continuum-discretized coupled-channels (CDCC) reaction framework \cite{Ra74b,YIK86,AIK87}. This method consists in discretizing the continuum making square-integrable functions, which guaranties that continuum-continuum couplings do not diverge. The continuum discretization is essentially performed in two ways: i) variational solutions of the projectile Hamiltonian are obtained at positive energies. Those are the pseudostates. ii) Continuum bins are constructed from averaging the scattering function over the wave number.  At higher energies, much above the Coulomb barrier, CDCC calculations could be time demanding, since they imply many partial waves. Therefore, an eikonal reaction framework is more suitable. This method relies on some simplifying assumptions at the high energy regime. Different versions and generalizations have been implemented \cite{BCG05,MBB02,AbS04,OY03,CBS08,PDB12,BCD09} since the original Glauber's publication \cite{Gl59}. In particular, the eikonal-CDCC method allows to study the influence of continuum states in reactions at high energies \cite{OY03}.

A microscopic continuum-discretized coupled-channels method (MCDCC) has been proposed in Ref. \cite{DH13}. In this reference the authors combine a microscopic cluster description of the projectile with the CDCC reaction framework. They applied the method to study the influence of continuum states on the elastic and inelastic scattering of an $^7$Li$=\alpha+t$ projectile, colliding with a non-composite $^{208}$Pb target at energies close to the Coulomb barrier. The aim of the present paper is to extend the study of Ref. \cite{DH13} of the elastic scattering of $^7$Li on lighter targets and at much higher energies. 

The method proposed in Ref. \cite{DH13} and followed in this work is expected to have a good predictive power since: i) it relies on microscopic wave functions of the projectile, which are calculated from effective nucleon-nucleon interactions. These wave functions reproduce experimental values as: ground state energies, electromagnetic transition probabilities, etc. ii) The method is based on nucleon-target interactions, instead of nucleus-nucleus interactions, which are available in a wide range of masses and energies. iii) There is no free parameter. 

The availability of CDCC elastic cross sections at energies higher than the Coulomb barrier, allows us to test the range of validity of the approximations relying on the multichannel eikonal method. Of course, the CDCC calculations are computational demanding, but they are exact, provide that convergence is reached, in the sense that no high energy approximations are made.

The paper is organized as follows. In Section \ref{theory} we describe the MCDCC method. Section \ref{results} is devoted to apply this method to describe the elastic scattering of $^7$Li on $^{12}$C and $^{28}$Si at $E_{\text{Lab}}=350$ MeV. We also illustrate $^7$Li$+^{12}$C inelastic and breakup angular distributions. In Section \ref{Meik-CDCC} we briefly describe the eikonal-CDCC approach and we incorporate a microscopic description of $^7$Li colliding with a non-composite $^{12}$C target. The high energy validity of these multichannel elastic cross sections is particularly tested in Section \ref{Test}.  Summary and conclusions are given in section \ref{conclusions}. 
\section{Microscopic CDCC method}\label{theory}
\subsection{Microscopic description of the projectile}
An intrinsic state of the projectile with angular momentum $J_\text{P}$, projection on $z$ $M_\text{P}$, and parity $\pi_\text{P}$, satisfies the Schr\"odinger equation
\begin{equation}
h_{\text{P}}\Psi_i^{J_\text{P}M_\text{P}\pi_\text{P}}(\xi_\text{P})=\epsilon_i^{J_\text{P}\pi_\text{P}}\Psi_i^{J_\text{P}M_\text{P}\pi_\text{P}}(\xi_\text{P}).
\label{micSch}
\end{equation}
Here $\xi_\text{P}=(\xi_1,\xi_2,\cdots,\xi_{A_\text{P}})$ notates the internal coordinates of the projectile, where $\xi_k$ includes the spatial, spin and isospin parts. The index $i$ labels bound states ($\epsilon_i^{J_\text{P}\pi_\text{P}}<0$) and pseudostates or variational solutions at positive energies ($\epsilon_i^{J_\text{P}\pi_\text{P}}>0$) of Eq. (\ref{micSch}). If we consider two-body interactions only, and that protons and neutrons in the projectile have approximately the same nucleon mass $M_N$, the projectile Hamiltonian $h_{\text{P}}$ is written as
\begin{equation}
h_{\text{P}}=\sum\limits_{k=1}^{A_\text{P}}T_k+\sum\limits_{k<j=1}^{A_\text{P}}V_{kj}-T_{c.m.}.
\label{micham}
\end{equation}

Let us take the projectile of mass $m_{\text{P}}=m_NA_{\text{P}}$ as made of two cluster nuclei with masses $m_1=m_NA_1$ and $m_2=m_NA_2$.  The resonating group method (RGM) \cite{Wh37} or its equivalent generator coordinate method (GCM) \cite{Ho77} provide variational solutions of the Schr\"odinger equation (\ref{micSch}). A GCM wave function is defined by
\begin{equation}
\Psi^{J_\text{P}M_\text{P}\pi_\text{P}}(\xi_\text{P})=\int d\vis S f^{J_\text{P}\pi_\text{P}}(\vis S)
\Upsilon^{J_\text{P}M_\text{P}\pi_\text{P}}(\vis S),
\label{GCMwf}
\end{equation}
where $\vis S$ is called generator coordinate and $\Upsilon^{J_\text{P}M_\text{P}\pi_\text{P}}(\vis S)$ is a basis function. 

We can construct a non-projected basis function $\Upsilon$ as the antisymmetrized product of two Slater determinants, each one associated with a cluster-nucleus, and constructed from harmonic oscillator shell model orbitals with oscillator parameter $B$. If all oscillator parameters of the single nucleon orbitals are equal, we can write this basis function as \cite{BR37}
\begin{equation}
\Upsilon (\vis S)=\phi_{cm}\mathcal{A}\phi^1(\varsigma_1)\phi^2(\varsigma_2)\Gamma(\vis \rho
-\vis S),
\label{basis}
\end{equation}
where $\mathcal{A}$ is the antisymmetrization operator, $\phi^i$ is the wave function of cluster $i$ with $\varsigma_i$ notating its set of translational invariant coordinates and $\phi_{cm}$ is the center of mass wave function of the projectile. 

The function $\Gamma$ is the shifted Gaussian function
\begin{equation}
\Gamma (\vis \rho -\vis S)=\left(\frac{\mu}{\pi B^2}\right)^{3/4}e^{-\mu (\vis \rho -\vis S)^2/2B^2},
\end{equation}
with $\vis \rho$ the relative coordinate between the center of mass of the clusters and $\mu=A_1A_2/A_{\text{P}}$.

Equation (\ref{basis}) is projected on angular momentum and parity \cite{Br66}. In practice, Eq. (\ref{GCMwf}) is discretized and the coefficients $f^{J_\text{P}\pi_\text{P}}(\vis S)$, after removing the center of mass, are determined variationally for bound states and pseudostates. 
\subsection{Projectile-Target Schr\"odinger equation}
Let us consider the scattering process of a composite projectile colliding with a non-composite target.
The total relative projectile-target Hamiltonian is written as
\begin{equation}
H(\vis R,\xi_\text{P})=-\frac{\hbar^2}{2\mu_{\text{PT}}}\nabla_R^2+h_\text{P}(\xi_\text{P})+V^{\text{PT}}(\vis R,\xi_\text{P}),
\label{priorH}
\end{equation}
where $\vis R$ is the relative coordinate between the center of mass of the projectile and the target. In spherical coordinates $\vis R=(\varphi_R,\theta_R,R)$, with $\varphi_R$ and $\theta_R$ the Polar and azimuthal angles, and $d\Omega_R=\sin\theta_Rd\theta_Rd\varphi_R$.

The first term on the right-hand side is the relative kinetic energy with reduced mass $\mu_{\text{PT}}=\frac{m_{\text{P}}m_{\text{T}}}{m_{\text{P}}+m_{\text{T}}}$, where $m_{\text{P}}$ and $m_{\text{T}}$ are the projectile and target masses. The last term is the projectile-target potential given by
\begin{equation}
V^{\text{PT}}(\vis R,\xi_\text{P})=\sum\limits_{k=1}^{A_\text{P}}V_{k\text{T}}(\vis R-\vis r_k),
\label{sumintpot}
\end{equation}
where $V_{k\text{T}}$ is the interaction of a nucleon $k$ in the projectile with the non-composite target. The position $\vis r_{k}$  of a nucleon $k$ in the projectile is defined from its  center of mass. 
 
A partial wave $\Phi^{JM\pi}(\vis R,\xi_\text{P})$ of total angular momentum $J$, with respective projection $M$ and parity $\pi$ satisfies the Schr\"odinger equation
\begin{equation}
H\Phi^{JM\pi}(\vis R,\xi_\text{P})=E_T\Phi^{JM\pi}(\vis R,\xi_\text{P}),
\label{schpost}
\end{equation}
with the total energy of the system given by
\begin{equation}
E_T=\epsilon_{i_0}^{J_{\text{P}_0}\pi_{\text{P}_0}}+E_{c.m.},
\label{totale}
\end{equation}
where $E_{c.m.}$ and $\epsilon_{i_0}^{J_{\text{P}_0}\pi_{\text{P}_0}}$ are the relative energy and internal energy of the projectile in the entrance channel. 
\subsection{CDCC coupled equations}
Let us expand the partial wave function $\Phi^{JM\pi}(\vis R,\xi_\text{P})$ as
\begin{equation}
\Phi^{JM\pi}(\vis R,\xi_\text{P})=\sum\limits_{iJ_\text{P}\pi_\text{P}L}\mathcal{Y}_{iJ_\text{P}\pi_\text{P}L}^{JM\pi}(\Omega_R,\xi_\text{P})\frac{\chi^{J}_{iJ_\text{P}\pi_\text{P}L}(R)}{R},
\label{cdccstate}
\end{equation}
with the basis functions
\begin{equation}
\mathcal{Y}_{iJ_\text{P}\pi_\text{P}L}^{JM\pi}(\Omega_R,\xi_\text{P})=\imath^L[Y_L\otimes\Psi^i_{J_\text{P}}]^{JM}.
\label{bfuncCDCC}
\end{equation}
Here $L$ is the relative orbital angular momentum of the projectile-target system.

The basis functions (\ref{bfuncCDCC}) satisfy the orthogonality relation
\begin{equation}
\left\langle\mathcal{Y}_{i'J'_\text{P}\pi'_\text{P}L'}^{JM'\pi}|\mathcal{Y}_{iJ_\text{P}\pi_\text{P}L}^{JM\pi}\right\rangle=
\delta_{i'i}\delta_{J'_\text{P}J_\text{P}}\delta_{\pi'_\text{P},\pi_\text{P}}\delta_{L'L}\delta_{M'M},
\end{equation}
where the Dirac notation indicates integration over $\Omega_R$ and the internal coordinates of the projectile. 

By inserting the state (\ref{cdccstate}) in the Schr\"odinger equation (\ref{schpost}) and projecting this equation on the functions (\ref{bfuncCDCC}), we end up with the set of coupled differential equations
\begin{multline}
\left[-\frac{\hbar^2}{2\mu_{\text{PT}}}\left(\frac{d^2}{dR^2}-\frac{L(L+1)}{R^2}\right)+\epsilon_{\alpha'}-E_T\right]\chi^{J}_{\alpha' L'}(R)\\
=-\sum\limits_{\alpha L}V_{\alpha'L',\alpha L}^{J\pi}(R)\chi^{J}_{\alpha L}(R),
\label{CDCC-eq}
\end{multline}
with $\alpha\equiv \{iJ_\text{P}\pi_\text{P}\}$. The diagonal and coupling potentials $V_{\alpha'L',\alpha L}^{J\pi}(R)$ are defined by (see the Appendix)
\begin{align}
V_{\alpha'L',\alpha L}^{J\pi}(R)&=\left\langle\mathcal{Y}_{\alpha'L'}^{JM\pi}|V^{\text{PT}}|\mathcal{Y}_{\alpha L}^{JM\pi}\right\rangle\notag\\
&=\sum\limits_{\lambda}C^{JJ_\text{P}J'_\text{P}}_{\lambda L L'}\langle\Psi_{i'}^{J'_\text{P}\pi'_\text{P}}||V_{\lambda}(R)||\Psi_i^{J_\text{P}\pi_\text{P}}\rangle,
\label{CDCC-cpots}
\end{align}
with the coefficients
\begin{align}
C^{JJ_\text{P}J'_\text{P}}_{\lambda L L'}=(-1)^{J'_\text{P}+J+\frac{1}{2}(L-L'+\lambda)}\frac{\hat{L}\hat{L'}\hat{\lambda}\hat{J'_\text{P}}}{\sqrt{4\pi}}\notag\\
\times
\begin{pmatrix}
L' & \lambda & L \\
0  & 0       & 0 \\
\end{pmatrix}
\begin{Bmatrix}
 J      & J'_\text{P} & L' \\
\lambda & L           & J_\text{P}
\end{Bmatrix},
\end{align}
where $\hat{x}=\sqrt{2x+1}$ and we have used the standard notations of the 3-J and 6-J symbols.

If we use GCM internal wave functions of the projectile, the reduced matrix element $\langle\Psi_{i'}^{J'_\text{P}\pi'_\text{P}}||V_{\lambda}(R)||\Psi_i^{J_\text{P}\pi_\text{P}}\rangle$ involves one-body matrix elements between Slater determinants, which can be determined systematically \cite{Br66}. An equivalent procedure is to employ a folding technique. In this case, the projectile-target interaction can be obtained by folding the nucleon-nucleus interactions with the microscopic densities of the projectile. In this context, the reduced matrix elements are given by \cite{SL79,KS00}
\begin{align}
&\langle\Psi_{i'}^{J'_\text{P}\pi'_\text{P}}||V_{\lambda}(R)||\Psi_i^{J_\text{P}\pi_\text{P}}\rangle\notag\\
&=\frac{1}{2\pi^2}\int_0^{\infty}dq~q^2 j_{\lambda}(qR)\left[\tilde{\rho}^{n(\lambda)}_{\alpha',\alpha}(q)\tilde{V}_{n\text{T}}(q)\right.\notag\\
&\hspace{4.cm}\left.+\tilde{\rho}^{p(\lambda)}_{\alpha',\alpha}(q)\tilde{V}_{p\text{T}}(q)\bigr.\right],
\label{reducedme}
\end{align}
where  $\tilde{\rho}^{n(\lambda)}_{\alpha',\alpha}(q)$ and $\tilde{\rho}^{p(\lambda)}_{\alpha',\alpha}(q)$ are Fourier multipoles of the diagonal and transition densities of the projectile. Note that the index $\alpha(\alpha')$ is associated with any bound state or pseudostate of the projectile. The terms $\tilde{V}_{n\text{T}}(q)$ and $\tilde{V}_{p\text{T}}(q)$ correspond to the Fourier transforms of central neutron-target and proton-target potentials. In the present work, bound and scattering states of $^7$Li are treated on the same footing and the GCM densities are determined following Ref. \cite{BDT94}.

The main ingredient in reaction calculations is the scattering matrix that allows to compute cross sections. This scattering matrix can be determined from the system of equations (\ref{CDCC-eq}), which can be solved by different methods \cite{NT99,IIL77,CH06,DBD10}. In particular, we use the R-matrix method on a Lagrange mesh \cite{DBD10}. It mainly consists in dividing the configuration space in two regions. An internal region, where each radial wave function $\chi^{J}_{\alpha L}(R)$ is expanded over a finite basis, and an external region, where each of these radial wave functions has reached its Coulomb asymptotic behavior. The matching of the wave function $\chi^{J}_{\alpha L}(R)$ at the boundary $a$  of both regions provides the collision matrix.

In practice, the sum in Eq. (\ref{CDCC-eq}) is truncated up to a maximal value of total angular momentum of the projectile $J_{\text{Pmax}}$ and the pseudostates are included up to determined excitation energy $E_{\text{max}}$. The contribution to the elastic cross sections beyond those values should be negligible.
\section{$^7$L\lowercase{i}$+^{12}$C and $^7$L\lowercase{i}$+^{28}$S\lowercase{i} elastic scattering with MCDCC}\label{results}
\subsection{Conditions of the calculations}
The calculations are essentially divided in two steps: i) computing the coupling potentials. ii) Determining the scattering matrix and cross sections. The coupling potentials have two main ingredients, the projectile bound and pseudostate wave functions, and the nucleon-target potentials.

The conditions to compute the $^7$Li wave functions are the same as in Ref. \cite{DH13}. The $^7$Li nucleus is described by an $\alpha+t$ cluster structure, and the Minnesota nucleon-nucleon interaction is used. This description provides a spectrum and a $B(E2,3/2^-\to 1/2^-)$ that are in good agreement with experiment. 

We consider central parts of optical potentials. The $n-^{12}$C and $p-^{12}$C interactions are taken from Ref. \cite{WP09} and the $n-^{28}$Si and $p-^{28}$Si interactions from Ref. \cite{KD03}. The multipole expansion of the potentials goes up to $\lambda_{\text{max}}=2$ in all cases. 

In order to determine the collision matrix we use the R-matrix method on a Lagrange-Legendre mesh of $N=130$ basis functions, with a channel radius $a=18$ fm. Several convergence tests were performed to check that beyond those values the cross sections do not vary at the scale of the figures. To compute the elastic cross sections, partial waves are summed up to a total angular momentum of the projectile-target system $J_{\text{max}}=200$.
\subsection{Elastic cross sections}
\begin{figure}[b]
\begin{center}
\includegraphics[scale=1]{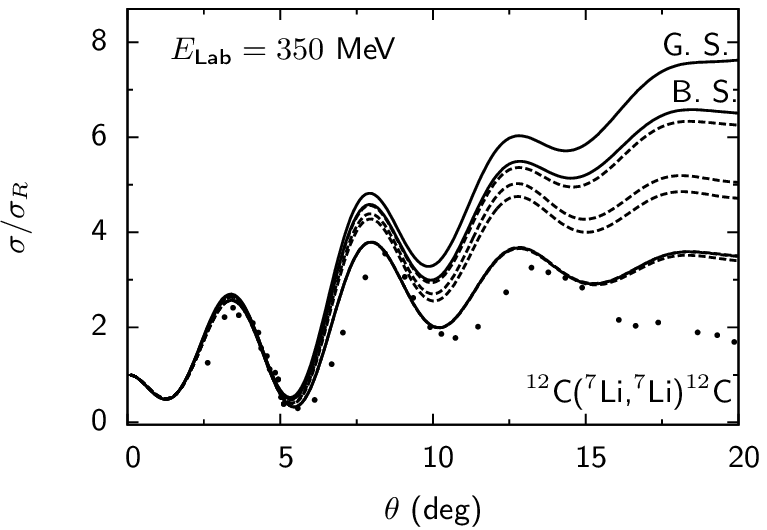}\\
\includegraphics[scale=1]{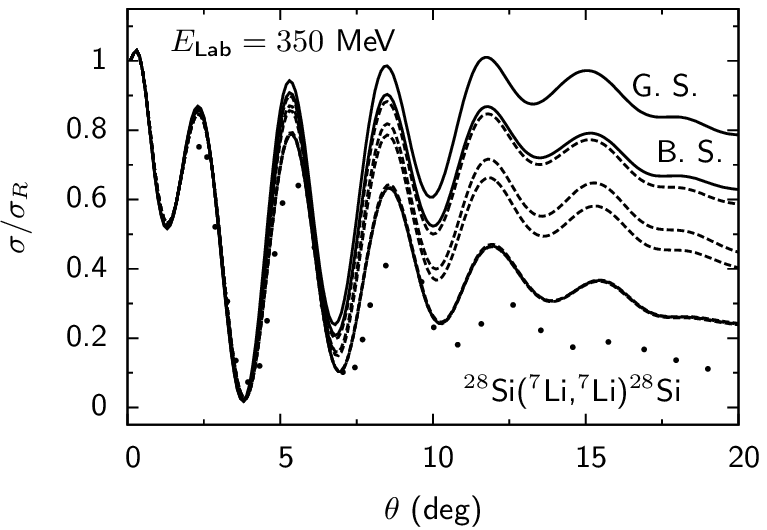}
\caption{CDCC Elastic cross sections (divided by the Rutherford cross section) of a composite $^7$Li impinging on $^{12}$C and $^{28}$Si targets. The solid lines labeled G.S. correspond to the single channel cross sections. The ones labeled B.S. include the $3/2^-$ and $1/2^-$ bound states only. The six dashed lines represent the calculations considering in addition to the bound states, the breakup channels up to a determined $J_{\text{Pmax}}$=1/2-11/2. Each $J_{\text{Pmax}}$ increases from the top to the bottom. The curves with $J_{\text{Pmax}}$=7/2-11/2 are superimposed. The points are the experimental data from Ref. \cite{Na95}.}
\label{CDCCconvj}
\end{center}
\end{figure}

In Fig. \ref{CDCCconvj} we display the CDCC elastic cross sections of $^7$Li on $^{12}$C and $^{28}$Si  at $E_\text{Lab}=350$ MeV. The cross sections are computed including various states of $^7$Li, where positive and negative parities are considered. 
We take into account the breakup channels up to a $J_{\text{Pmax}}=11/2$ and a cutoff excitation energy $E_{\text{max}}=15$ MeV, defined from the $\alpha+t$ threshold. 
The calculations for both targets converge at $J_{\text{Pmax}}=7/2$, which can be understood, since the maximum spin of the well known cluster state resonances is $7/2$ ($J^{\pi_{\text{P}}}_{\text{P}}=7/2^-$, $E_{\text{res}}\simeq 2.18$ MeV).

Figure \ref{CDCCconvj} shows a strong influence on the elastic cross sections of the excited state and of the breakup channels at approximately $\theta > 7^\circ$ for both nuclei. The converged cross sections are in very good agreement with the experimental data up to $15^\circ$ for $^{12}$C and $10^\circ$ for $^{28}$Si. At larger angles, our predictions overestimate the data about a factor of $2$ for both systems. Qualitatively, the CDCC calculations in the whole angular range studied, are good predictions since there is no free parameter.

\begin{figure}[t]
\begin{center}
\includegraphics[scale=1.05]{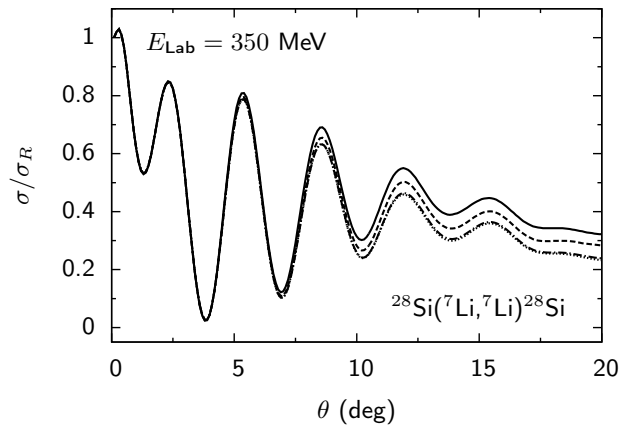}
\caption{Convergence of the CDCC elastic scattering cross section with the cutoff excitation energy of the projectile $E_{\text{max}}$. The solid, dashed, dashed-dotted and dotted lines correspond to $E_{\text{max}}=$7 MeV, 11 MeV, 15 MeV and 19 MeV respectively. The calculations include breakup channels up to $J_{\text{Pmax}}=7/2$. The curve corresponding to $E_{\text{max}}=15$ MeV is superimposed with the curve $E_{\text{max}}=19$ MeV.}
\label{conve28Si}
\end{center}
\end{figure}

\begin{figure}[b]
\begin{center}
\includegraphics[scale=1.05]{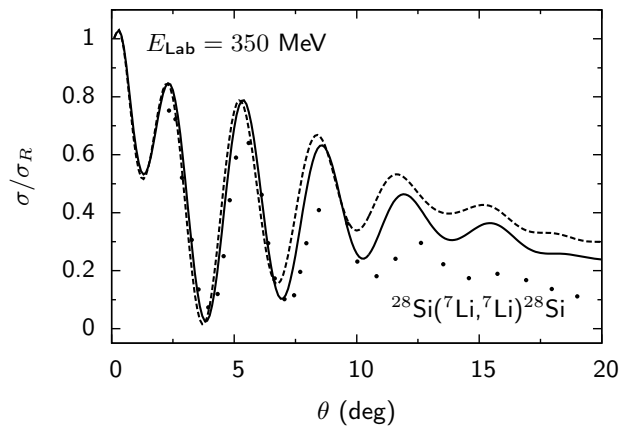}
\caption{Influence of the nucleon-target nuclear potential. The solid and dashed lines correspond to the Koning and Delaroche \cite{KD03} and  Weppner \textit{et al.} \cite{WP09} optical potentials, respectively.}
\label{Pot-inf}
\end{center}
\end{figure}

In Fig. \ref{conve28Si} is illustrated the $^7$Li$+^{28}$Si elastic cross sections when $E_{\text{max}}$ is progressively increased. We see that the convergence is reached at $E_{\text{max}}\approx 15$ MeV. A similar convergence behavior is obtained for the $^{12}$C target and  it is therefore not shown.

On the other hand, as the $^7$Li elastic scattering is on light targets, the process is nuclear dominated and the behavior at large angles is strongly influenced by the nuclear contribution. Thus, we study the influence of the choice of the nuclear nucleon-target potential. Fig. \ref{Pot-inf} compares the $^7$Li$+^{28}$Si elastic scattering using two nuclear optical potentials. The potentials of Koning and Delaroche, employed to computed the curves in Figs. \ref{CDCCconvj} and \ref{conve28Si}, and of Weppner \textit{et al.} \cite{WP09}. Both have similar Woods-Saxon functional forms in their volume and surface terms. In the angular range shown, they reproduce successfully the $n-^{28}$Si and $p-^{28}$Si experimental elastic cross sections at energies close to 50 MeV (see Ref. \cite{WP09}). 

Figure \ref{Pot-inf} shows that the cross sections are slightly affected by the choice of the potential at large angles ($\theta>7^\circ$) with a difference around $20\%$ between them.
\subsection{Inelastic and breakup cross sections}
\begin{figure}[b]
\begin{center}
\includegraphics[scale=1.05]{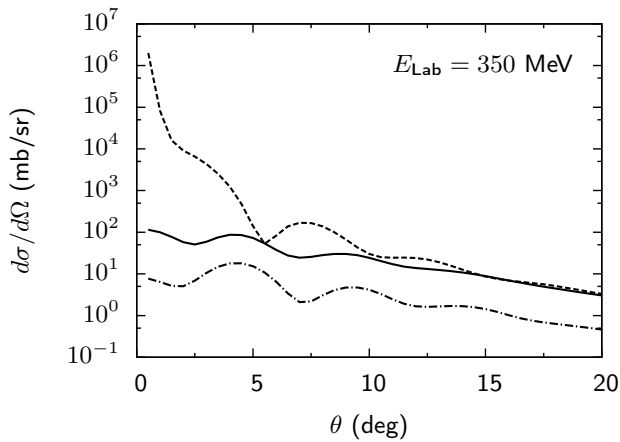}
\caption{CDCC $^7$Li$+^{12}$C breakup (solid line), elastic (dashed line) and inelastic (dashed-dotted line) angular distributions.}
\label{comp-reac}
\end{center}
\end{figure}

In Fig. \ref{comp-reac} we illustrate the MCDCC predictions of the inelastic and breakup angular distributions of $^7$Li$+^{12}$C system at $E_\text{Lab}=350$ MeV, compared with the converged elastic cross section shown in Fig. \ref{CDCCconvj}.
The breakup angular distribution is estimated by summing over all the individual inelastic excitations from the bound state to a pseudostate. As expected, the elastic scattering dominates at small angles $\theta<5^\circ$. At $\theta>10^\circ$ the elastic and breakup cross sections are very close to each other, in correspondence with the range where the breakup channels influence the elastic cross section the most (see Fig. \ref{CDCCconvj}). The inelastic process is the less likely to occur in the whole angular range. 
\section{Microscopic Eikonal-CDCC elastic scattering}\label{Meik-CDCC}
Combining precise projectile wave functions that consider the internal structure of at least one of the collision partners, should provide a suitable framework of nuclear reaction studies. However, it increases the complication level both theoretically and computationally, especially at high energies when the number of partial waves involved increases. The approximations relying on the eikonal method make it simpler and less computing demanding than the CDCC calculations. 

Some works that utilize in eikonal methods microscopic cluster wave functions of the projectile and nucleon-target scattering information, have been introduced to describe nucleus-nucleus reactions \cite{AS00a,AS00b,PD10a}. In particular, we investigate in Ref. \cite{PD10a} the elastic scattering of an alpha projectile at high energies, using a GCM wave function. This wave function is a four-nucleon Slater determinant corresponding to a single cluster approximation. As the $^4$He nucleus is in the ground state, there is no need of angular projection, which is one of the main issues in multi-cluster microscopic calculations. 

In the present work, we use a more complicated projectile than in Ref. \cite{PD10a}, $^7$Li, and a multichannel framework. The microscopic projectile is impinging on a $^{12}$C target at 350 MeV. To this end, we use the eikonal-CDCC method proposed in Ref. \cite{OY03} and we incorporate a microscopic description into it. 

The eikonal-CDCC method is based on solving the system of first order differential equations (see Ref. \cite{OY03} for details)
\begin{multline}
\frac{i\hbar^2 K_{c}}{\mu_{\text{PT}}}\frac{\partial}{\partial Z}\psi^{\alpha_0 M_{\text{P}_0}}_{\alpha M_\text{P}}(b,Z)=\\
\sum\limits_{\alpha' M'_\text{P}}e^{i(K_{c'}-K_{c})Z}V_{\alpha M_\text{P},\alpha' M'_\text{P}}(b,Z)\psi^{\alpha_0 M_{\text{P}_0}}_{\alpha' M'_\text{P}}(b,Z),
\label{eik-ceq}
\end{multline}
where we use  $\vis R$ in cylindrical coordinates i.e. $\vis R=(\vis b,Z)$,  with $\vis b$ the transverse component of $\vis R$. In Eq. (\ref{eik-ceq}) the superscripts indicate the entrance channel with the set $\{\alpha_0 M_{\text{P}_0}\}$ and the wave number $K_c$ is defined by
\begin{equation}
\frac{\hbar^2}{2\mu_{\text{PT}}}K^{2}_{c}=E_T-\epsilon_{\alpha}-\frac{\hbar^2}{2\mu_{\text{PT}}}\frac{(M_{\text{P}}-M_{\text{P}_0})^2}{b^2}.
\label{wnumberb}
\end{equation}

Equation (\ref{eik-ceq}) is derived after ignoring the kinetic energy term in a CDCC-like system of coupled differential equations. This condition is valid at incident energies much higher than the Coulomb barrier. System (\ref{eik-ceq}) is solved with the initial condition
\begin{equation}
\psi^{\alpha_0 M_{\text{P}_0}}_{\alpha M_\text{P}}(b,Z\to-\infty)=\delta_{\alpha\alpha_0}\delta_{M_\text{P}M_{\text{P}_0}}.
\label{initialc}
\end{equation}
This condition means that there is a bound state multiplied by a plane wave in the entrance channel. 

The eikonal-CDCC diagonal and coupling potentials are given by
\begin{align}
&V_{\alpha' M'_\text{P},\alpha M_\text{P}}(\vis R)\notag\\
&\hspace{1.3cm}=\langle\Psi_{i'}^{J'_\text{P}M'_\text{P}\pi'_\text{P}}|V^{\text{PT}}|\Psi_i^{J_\text{P}M_\text{P}\pi_\text{P}}\rangle,\notag\\
&\hspace{1.3cm}=e^{i(M_\text{P}-M'_\text{P})\varphi_R}
V_{\alpha' M'_\text{P},\alpha M_\text{P}}(\theta_R,R).
\label{intpot}
\end{align}
The Dirac notation stands for integration over the internal coordinates of the projectile only. The term $V_{\alpha' M'_\text{P},\alpha M_\text{P}}(\theta_R,R)$ is given by (see the Appendix)
\begin{multline}
V_{\alpha' M'_\text{P},\alpha M_\text{P}}(\theta_R,R)
=\sum\limits_{\lambda}\tilde{C}^{J_\text{P}J'_\text{P}\lambda}_{M_\text{P}M'_\text{P}}
P_{\lambda}^{M'_\text{P}-M_\text{P}}(\cos\theta_R)\\
\times\langle\Psi_{i'}^{J'_\text{P}\pi'_\text{P}}||V_{\lambda}(R)||\Psi_i^{J_\text{P}\pi_\text{P}}\rangle,
\label{cpot}
\end{multline}
where the coefficients $\tilde{C}^{J_\text{P}J'_\text{P}\lambda}_{M_\text{P}M'_\text{P}}$ are defined as
\begin{align}
\tilde{C}^{J_\text{P}J'_\text{P}\lambda}_{M_\text{P}M'_\text{P}}=&(-i)^{\lambda}\sqrt{\frac{(2\lambda+1)(\lambda-M'_\text{P}+M_\text{P})!}{4\pi(\lambda+M'_\text{P}-M_\text{P})!}}\notag\\
&\times(J_\text{P}M_\text{P}\lambda M'_\text{P}-M_\text{P}|J'_\text{P}M'_\text{P}).
\end{align}
The reduced matrix elements in Eq. (\ref{cpot}) are common to the CDCC potentials of Sec. \ref{theory} (Eq. (\ref{CDCC-cpots})). They are determined from Eq. (\ref{reducedme}).

The elastic angular distribution is calculated from \cite{CH13}
\begin{equation}
\frac{d\sigma}{d\Omega}=\frac{1}{2J_{\text{P}_0}+1}\sum\limits_{M'_{\text{P}_0}M_{\text{P}_0}}|f_{M'_{\text{P}_0}M_{\text{P}_0}}(\theta)|^2,
\end{equation}
with the elastic scattering amplitude 
\begin{equation}
f_{M'_{\text{P}_0}M_{\text{P}_0}}(\theta)=-2\pi^2\left(\frac{2\mu_{\text{PT}}}{\hbar^2}\right)T_{M'_{\text{P}_0}M_{\text{P}_0}}(\theta)
\end{equation}
and the transition matrix element $T_{M'_{\text{P}_0}M_{\text{P}_0}}$ \cite{OY03}
\begin{align}
T_{M'_{\text{P}_0}M_{\text{P}_0}}\simeq&\frac{i^{-\nu+1}\hbar^2}{(2\pi)^{2}\mu_{\text{PT}}}\notag\\
&\times\int_{0}^{\infty}dbb K_cJ_{\nu}(K_cb\theta)\left(S_{M'_{\text{P}_0},M_{\text{P}_0}}(b)\right.\notag\\
&\hspace{3.7cm}\left.-\delta_{M'_{\text{P}_0}M_{\text{P}_0}}\right),
\label{tmeeik}
\end{align}
where $J_\nu$ is the Bessel function of the first kind of order $\nu=|M'_{\text{P}_0}-M_{\text{P}_0}|$, and we use the definition of the eikonal scattering amplitudes
\begin{equation}
S_{M'_{\text{P}_0},M_{\text{P}_0}}(b)=\psi^{M_{\text{P}_0}}_{M'_{\text{P}_0}}(b,Z\to\infty).
\label{seikamp}
\end{equation}
Those are obtained by integrating over $Z$ the system (\ref{eik-ceq}) with a fourth-order Runge-Kutta method. For the $^7$Li$+^{12}$C elastic scattering, we integrate the equations with a step $h_z=0.1$ fm up to 30 fm. The integral over $b$ given in Eq. (\ref{tmeeik}) is performed up to 30 fm with a step $h_b=0.1$ fm. 

The calculation of the scattering matrix elements is from a $b_{\text{min}}$ to avoid imaginary values of $K_{c}$ in Eq. (\ref{wnumberb}). There is almost no effect of this approximation on the elastic cross section, since the elastic scattering matrix elements $S_{M'_{\text{P}_0},M_{\text{P}_0}}(b)$ for $b<b_{\text{min}}$ are negligible in comparison with those at higher $b$ values (typical $b_{\text{min}}<$ 1 fm). As usual, to compute the cross sections we separate the Coulomb projectile-target eikonal scattering amplitudes to get faster convergence \cite{BCD09}.

\begin{figure}[t]
\begin{center}
\includegraphics[scale=1.05]{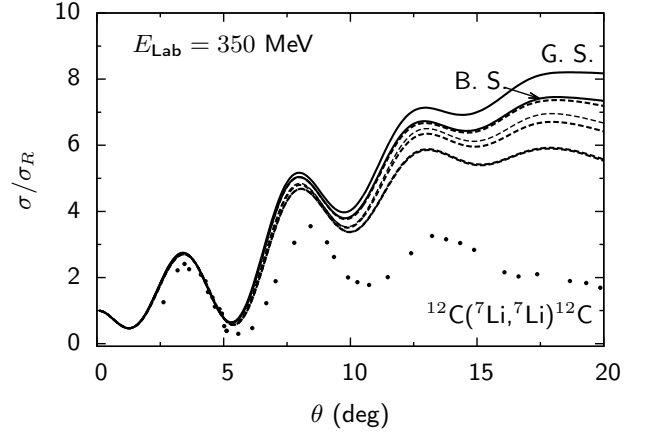}
\caption{$^7$Li$+^{12}$C eikonal-CDCC elastic cross section (divided by the Rutherford cross section). The solid lines labeled G.S. correspond to the single channel cross sections. The ones labeled B.S. include the $3/2^-$ and $1/2^-$ bound states only. The six dashed lines represent the calculations considering in addition to the bound states, the breakup channels up to a determined $J_{\text{Pmax}}$=1/2-11/2. Each $J_{\text{Pmax}}$ increases from the top to the bottom. The curves with $J_{\text{Pmax}}$=7/2-11/2 are superimposed. The points are the experimental data from Ref. \cite{Na95}.}
\label{convj12C}
\end{center}
\end{figure}

Figure \ref{convj12C} shows the single channel (GS) and multichannel calculations. At large angles ($\theta>7^\circ$), we observe an influence of the breakup channels on the elastic scattering, but the agreement with the experimental data is poor. In contrast, the CDCC cross section shown in Fig. \ref{CDCCconvj}  describes much better the experimental data indicating that the collision energy is not high enough for the high energy approximation relying on the eikonal-CDCC approach to be valid. This argument is discussed in detail in the next section.
\section{Test of the eikonal approach}\label{Test}
As we are able to calculate CDCC elastic cross sections at energies much higher than the Coulomb barrier, 
\begin{figure}[htb]
\begin{center}
\includegraphics[scale=0.95]{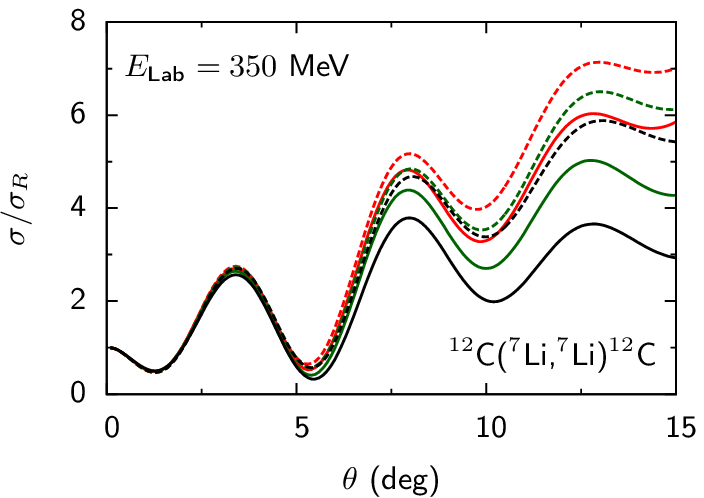}
\includegraphics[scale=0.95]{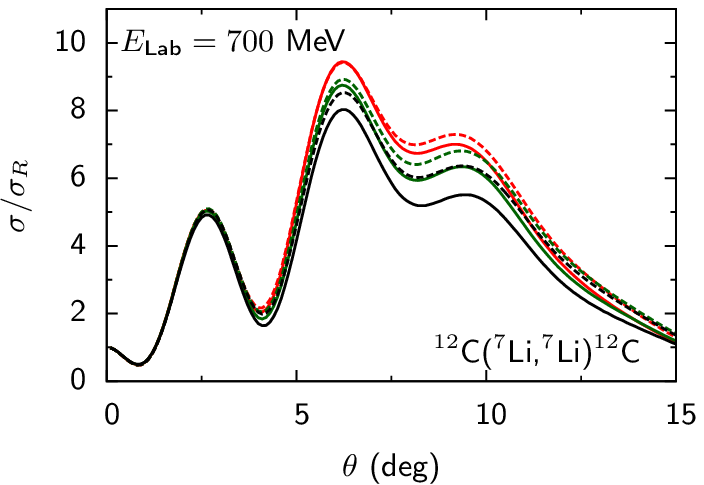}
\includegraphics[scale=0.95]{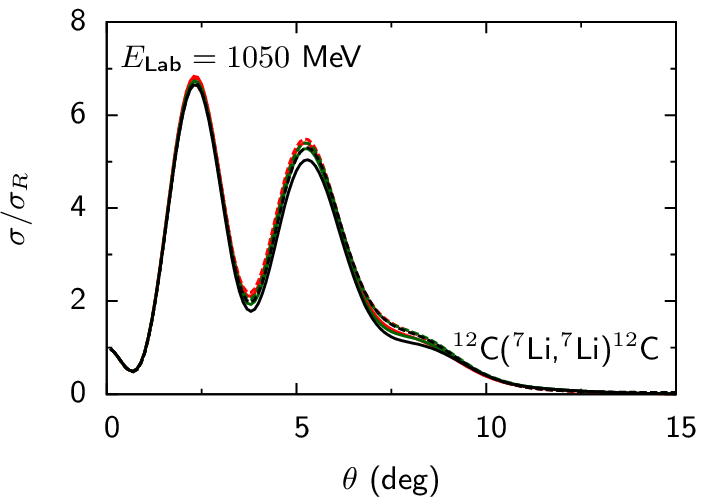}
\caption{(Color online) Comparison between the CDCC (solid lines) and eikonal-CDCC (dashed lines) elastic cross sections at different incident energies. The red curves correspond to single channel predictions. The green (black) curves are the calculations including in addition to the bound states the breakup channels up to $J_{\text{Pmax}}=3/2$ ($J_{\text{Pmax}}=7/2$).}
\label{eikcdcc12C}
\end{center}
\end{figure}
let us take advantage of this fact and consider those as reference calculations to test the high energy range of validity of multichannel eikonal cross sections. 

We consider the $^7$Li$+^{12}$C system and in addition to $E_{\text{Lab}}=350$ MeV, we perform calculations at 700 MeV and 1050 MeV. The conditions to compute the eikonal-CDCC cross sections are the same as described for $E_{\text{Lab}}=350$ MeV, with smaller $h_b$ of 0.05 fm at 700 MeV and 0.01 fm at 1050 MeV. The CDCC cross sections are obtained with $N=150$ and $J_{\text{max}}=300$ at 700 MeV, and with $N=180$ and $J_{\text{max}}=550$ at 1050 MeV. For both cases we use $a=18$ fm.

The comparison between the CDCC and eikonal-CDCC cross sections is displayed in Fig. \ref{eikcdcc12C}. We include three different kind of calculations: single channel, breakup channels up to $J_{\text{Pmax}}=3/2$ and breakup channels up to $J_{\text{Pmax}}=7/2$. 

At $E_{\text{Lab}}=350$ MeV, we observe a large difference between the eikonal-CDCC and CDCC calculations at $\theta> 7^\circ$. In addition, the influence of the breakup channels on the elastic eikonal-CDCC cross sections is much reduced than in the CDCC predictions. 

We notice that the agreement between both kind of methods improves at $E_{\text{Lab}}=700$ MeV. Increasing the energy up to $E_{\text{Lab}}=1050$ MeV reduces the difference between them to less than 0.15. However, the influence of the breakup channels on the elastic cross sections becomes very small, which can be understood since the projectile excitation energies are much smaller than the incident energy and therefore we can neglect the projectile inner motion. Indeed, this is the main idea behind adiabatic models \cite{TN09}. Similar behavior is presented in the CDCC elastic scattering scattering of $^{11}$Be+$^{64}$Zn \cite{DD12}.
\section{Summary and conclusions}\label{conclusions}
We use a MCDCC reaction framework \cite{DH13} to describe elastic scattering at high energies. This method is applied to the $^7$Li$=\alpha+t$ nucleus impinging on $^{12}$C and $^{28}$Si targets. The $^7$Li nucleus is weakly bound and coupling with the continuum is expected to play an important role in the description of the elastic cross section \cite{DH13}. We calculate the projectile-target interactions by using a folding technique, where the main ingredients are the $^7$Li GCM wave functions and the nucleon-target potentials. The CDCC scattering matrix is obtained from solving the CDCC system of coupled equations through the R-matrix method on a Lagrange mesh. 

First, we compute CDCC elastic cross sections at $E_{\text{Lab}}=350$ MeV, where experimental data is available. We observe an influence of the breakup channels on the elastic cross sections. These breakup channels have to be included in order to get a better description of the experimental data at large angles ($\theta>7^\circ$). The improvement, with respect to the single channel results, is around a factor of 2 for $^{12}$C, and 4 for $^{28}$Si.

Next, we study the influence of the nucleon-target nuclear potential on the $^7$Li$+^{28}$Si cross section. To this end we use two nucleon-$^{28}$Si potentials and obtain a sensitivity around $20\%$ at $\theta> 7^\circ$. The influence of the choice of the nucleon-nucleon interaction used to compute the projectile densities, should be addressed in future works.

On the other hand, we compute the eikonal-CDCC elastic cross section of a microscopic $^7$Li$=\alpha+t$ impinging on a non-composite $^{12}$C at $E_{\text{Lab}}=350$ MeV. The eikonal-CDCC result deviates significantly from the CDCC one, which is closer to the experimental data. The disagreement is explained as the high energy validity relying on the multichannel eikonal treatment is not satisfied. Thus, in order to test multichannel eikonal elastic cross sections, we compare the eikonal-CDCC and CDCC elastic cross sections at different incident energies for the $^{12}$C target, taking as reference the CDCC calculations.  Increasing the incident energy improves the agreement between the eikonal-CDCC and CDCC calculations showing that for the multichannel eikonal cross sections be fairly valid in the whole angular range shown, $E_{\text{Lab}}$ must be at least $\sim 1000$ MeV. Even though, at such energy, the contribution of the breakup channels becomes very small.

The present work represents an improved perspective in nucleus-nucleus scattering at high energies. It can be extended to other exotic nuclei, as Borromean nuclei, or to other reactions such inelastic scattering, breakup or fusion.

\begin{acknowledgments}
This text presents research results of the IAP programme P7/11 initiated by the Belgian-state 
Federal Services for Scientific, Technical and Cultural Affairs. 
E.C.P. is supported by the IAP programme. P.D. acknowledges the support of F.R.S.-FNRS, Belgium. 
\end{acknowledgments}

\appendix*
\section{Interaction potential}\label{App_pot}
\subsection{Diagonal and coupling potentials used in the CDCC equations}
Let us expand the projectile-target potential in multipoles as
\begin{eqnarray}
V^{\text{PT}}(\vis R,\xi_\text{P})=\sum\limits_{\lambda}(-i)^{\lambda}Y_{\lambda}^{*M'_\text{P}-M_\text{P}}(\Omega_R)V_{\lambda}(R,\xi_\text{P}).\nonumber\\
\label{pwtpot}
\end{eqnarray}
This potential can be rewritten as the sum of products of multipole tensor operators
\begin{equation}
V^{\text{PT}}=\sum\limits_{\lambda}i^{-\lambda}(Y_{\lambda}\cdot V_{\lambda}).
\label{pottensor}
\end{equation}
By using the definition (\ref{pottensor}), the  potentials defined in Eq. (\ref{CDCC-cpots}) become
\begin{align}
&V_{\alpha'L',\alpha L}^{J\pi}(R)\notag\\
&=\left\langle\mathcal{Y}_{\alpha' L'}^{JM\pi}|V^{\text{PT}}|\mathcal{Y}_{\alpha L}^{JM\pi}\right\rangle\notag\\
&=\sum\limits_{\lambda} i^{L-L'-\lambda}\left \langle [Y_{L'}\otimes\Psi^{i'}_{J'_\text{P}}]^{JM}|Y_{\lambda}\cdot V_{\lambda}|[Y_L\otimes\Psi^i_{J_\text{P}}]^{JM}\right\rangle,
\label{cpotCDCC}
\end{align}
where the Dirac notation represents integration over $\Omega_R$ and the projectile internal coordinates.

If we use the Wigner-Eckart's theorem \cite{Ed57} in expression (\ref{cpotCDCC}) we end up with
\begin{align}
&V_{\alpha'L',\alpha L}^{J\pi}(R)
&=\sum\limits_{\lambda}C^{J_\text{P}J'_\text{P}\lambda}_{J L L'}\langle\Psi_{i'}^{J'_\text{P}\pi'_\text{P}}||V_{\lambda}(R)||\Psi_i^{J_\text{P}\pi_\text{P}}\rangle,
\end{align}
where
\begin{align}
C^{JJ_\text{P}J'_\text{P}}_{\lambda L L'}=(-1)^{J'_\text{P}+J+\frac{1}{2}(L-L'+\lambda)}\frac{\hat{L}\hat{L'}\hat{\lambda}\hat{J'_\text{P}}}{\sqrt{4\pi}}\notag\\
\times
\begin{pmatrix}
L' & \lambda & L \\
0  & 0       & 0 \\
\end{pmatrix}
\begin{Bmatrix}
 J      & J'_\text{P} & L' \\
\lambda & L           & J_\text{P}
\end{Bmatrix}
.
\end{align}
\subsection{Diagonal and coupling potentials used in the eikonal-CDCC equations}
Let us define the diagonal and coupling potentials depending on $\vis R$ by
\begin{equation}
V_{\alpha'M'_{\text{P}},\alpha M_{\text{P}}}(\vis R)
=\langle\Psi_{i'}^{J'_\text{P}M'_\text{P}\pi'_\text{P}}|V^{\text{PT}}|\Psi_i^{J_\text{P}M_\text{P}\pi_\text{P}}\rangle,
\label{intpot2}
\end{equation}
where the Dirac notation stands for integration over the internal coordinates of the projectile. If we introduce the expansion (\ref{pwtpot}) in Eq. (\ref{intpot2}) we get
\begin{align}
&V_{\alpha'M'_{\text{P}},\alpha M_{\text{P}}}(\vis R)\notag\\
&\hspace{0.5cm}=\sum\limits_{\lambda}(-i)^{\lambda}(J_\text{P}M_\text{P}\lambda M'_\text{P}-M_\text{P}|J'_\text{P}M'_\text{P})\notag\\
&\hspace{1.5cm}\times Y_{\lambda}^{*M'_\text{P}-M_\text{P}}(\Omega_R)\langle\Psi_{i'}^{J'_\text{P}\pi'_\text{P}}||V_{\lambda}(R)||\Psi_{i}^{J_\text{P}\pi_\text{P}}\rangle.
\label{VPT}
\end{align}
Here we have used the Wigner-Eckart's theorem. 

By expressing the spherical Harmonics in terms of the associate Legendre polynomials we have
\begin{align}
&V_{\alpha'M'_{\text{P}},\alpha M_{\text{P}}}(\vis R)\notag\\
&\hspace{1.3cm}=\langle\Psi_{i'}^{J'_\text{P}M'_\text{P}\pi'_\text{P}}|V^{\text{PT}}|\Psi_{i}^{J_\text{P}M_\text{P}\pi_\text{P}}\rangle,\notag\\
&\hspace{1.3cm}=e^{i(M_\text{P}-M'_\text{P})\varphi_R}
V_{\alpha'M'_{\text{P}},\alpha M_{\text{P}}}(\theta_R,R),
\label{ipot}
\end{align}
with
\begin{align}
V_{\alpha'M'_{\text{P}},\alpha M_{\text{P}}}&(\theta_R,R)\notag\\
&=\sum\limits_{\lambda}\tilde{C}^{J_\text{P}J'_\text{P}J}_{M_\text{P}M'_\text{P}}
P_{\lambda}^{M'_\text{P}-M_\text{P}}(\cos\theta_R)\notag\\
&\hspace{1.3cm}\times\langle\Psi_{i'}^{J'_\text{P}\pi'_\text{P}}||V_{\lambda}(R)||\Psi_i^{J_\text{P}\pi_\text{P}}\rangle,
\label{CCpot}
\end{align}
and
\begin{align}
\tilde{C}^{J_\text{P}J'_\text{P}J}_{M_\text{P}M'_\text{P}}=&(-i)^{\lambda}\sqrt{\frac{(2\lambda+1)(\lambda-M'_\text{P}+M_\text{P})!}{4\pi(\lambda+M'_\text{P}-M_\text{P})!}}\notag\\
&\times(J_\text{P}M_\text{P}\lambda M'_\text{P}-M_\text{P}|J'_\text{P}M'_\text{P}).
\end{align}

\bibliographystyle{apsrev4-1}
\bibliography{biblio2}
\end{document}